\documentclass[%
preprint,
showpacs,reprintnumbers,
 amsmath,amssymb,
 aps,
prl,
superscriptaddress,
showpacs, showkeys]{revtex4-1}
\usepackage{amsmath}
\usepackage{subfigure}
\usepackage{textgreek}
\usepackage{color}
\usepackage{breqn}
\usepackage{graphicx}
\usepackage{dcolumn}
\usepackage{bm}
\usepackage{hyperref}
\usepackage{textcomp}
\usepackage{xr}
\hypersetup{bookmarksnumbered, pdfpagemode=UseOutlines, 
colorlinks=true, citecolor=blue, filecolor=blue, linkcolor=blue, urlcolor=blue}

 \makeatletter
 \let\cat@comma@active\@empty
 \makeatother

\graphicspath{/home/sid/Dropbox/6th_paper_WS2_susbstrate/paper_fig
}

\begin{document}

\preprint{}

\title{Spin transport  in high-mobility graphene on WS$_{\text{2}}$ substrate with electric-field tunable proximity spin-orbit interaction}

\author{S. Omar} 
\thanks{corresponding author}
\email{s.omar@rug.nl}
\affiliation{The Zernike Institute for Advanced Materials University of Groningen Nijenborgh 4 9747 AG, Groningen, The Netherlands}
\author{B.J. van Wees}
\affiliation{The Zernike Institute for Advanced Materials University of Groningen Nijenborgh 4 9747 AG, Groningen, The Netherlands}%
\date{\today}

\begin{abstract}
Graphene supported on a transition metal dichalcogenide substrate offers a novel platform to study the spin transport in graphene in presence of a substrate induced spin-orbit coupling, while preserving its intrinsic charge transport properties.
We report the first non-local spin transport measurements in graphene completely supported on a 3.5 nm thick tungsten disulfide (WS$_{\text{2}}$) substrate, and encapsulated from the top with a 8 nm thick hexagonal boron nitride layer. For graphene, having mobility up to 16,000 cm$^2$V$^{-1}$s$^{-1}$, we measure almost constant spin-signals both in electron and hole-doped regimes, independent of the conducting state of the underlying WS$_2$ substrate, which rules out the role of spin-absorption by WS$_2$. The spin-relaxation time $\tau_{\text{s}}$ for the electrons in graphene-on-WS$_2$ is drastically reduced down to $\sim$ 10 ps than $\tau_{\text{s}} \sim$  800 ps in graphene-on-SiO$_2$ on the same chip. The strong suppression of $\tau_{\text{s}}$ along with a detectable weak anti-localization signature in the quantum magneto-resistance  measurements is a clear effect of the WS$_2$  induced spin-orbit coupling (SOC) in graphene. Via the top-gate voltage application in the encapsulated region, we modulate the electric field by 1 V/nm, changing $\tau_{\text{s}}$ almost by a factor of four which suggests the electric-field control of the in-plane Rashba SOC. Further, via carrier-density dependence of $\tau_{\text{s}}$ we also identify the fingerprints of the D'yakonov-Perel' type mechanism in the hole-doped regime at the graphene-WS$_2$ interface.
\end{abstract}

\keywords{Spintronics, Graphene, graphene-semiconductor interface, spin-orbit coupling}
\maketitle

\section{introduction}

Recent exploration of various two-dimensional (2D) materials and their heterostructures has provided access to novel charge \cite{giovannetti_substrate-induced_2007,woods_commensurate-incommensurate_2014} and spin-related phenomena \cite{yang_strong_2017,yang_tunable_2016,wang_strong_2015,wang_origin_2016, omar_graphene-$mathrmws_2$_2017, gurram_bias_2017} which are either missing or do not have a measurable effect in intrinsic graphene.  Graphene (Gr) can interact with the neighboring material via weak van der Waals interactions which help to preserve its intrinsic charge transport properties while it can still acquire some foreign properties from the host substrate such as a sizable band gap in Gr-on-hexagonal Boron Nitride (hBN) substrate at the Dirac point due to a sublattice dependent crystal potential in graphene \cite{giovannetti_substrate-induced_2007,woods_commensurate-incommensurate_2014}. For  Gr-transition metal dichalcogenide (TMD) heterostructures, an enhanced intrinsic spin-orbit coupling (SOC) in the order of 5-15 meV can be induced in graphene, along with a meV order valley-Zeeman splitting due to inequivalent K and K' valleys in graphene \cite{wang_origin_2016,gmitra_graphene_2015}, a Rashba SOC due to breaking of the inversion symmetry at the graphene-TMD interface \cite{yang_strong_2017,yang_tunable_2016} with a possibility of spin-valley coupling \cite{xiao_valley-contrasting_2007,cummings_giant_2017}. This unique ability of the graphene-TMD interface makes it an attractive platform for studying the spin-related proximity induced effects in graphene. 
 
In recent reports of spin-transport in graphene-TMD heterostructures \cite{dankert_electrical_2017, yan_two-dimensional_2016}, a reduced spin-signal and spin-relaxation time were measured in graphene when the TMD was in conducting state. This behavior was attributed to the spin-absorption/enhanced SOC via the TMD. On the contrary, in weak anti-localization (WAL) magnetotransport measurements \cite{wang_origin_2016, wang_strong_2015}, a reduced spin-relaxation time, independent of the carrier-type, carrier-density in graphene and the conducting state of the TMD was observed which was attributed to a greatly enhanced SOC in graphene via the proximity effect of the TMD. Also, the existence of the interplay between the valley-Zeeman and Rashba SOC was  theoretically \cite{cummings_giant_2017} and experimentally \cite{ghiasi_large_2017,benitez_strongly_2017} demonstrated in the anisotropy of the spin relaxation time for the out-of-plane and in-plane spin-signals in TMD-graphene heterostructures. 

Surrounded by distinct conclusions, which seem to depend on the device geometry and experiment-type, it  calls for revisiting the problem in a different way, i.e., a  direct spin-transport measurement using TMD as a substrate for graphene. 
It has multiple advantages: i) similar to hBN, TMD substrates have already shown significantly improved charge transport properties of graphene \cite{kretinin_electronic_2014} than graphene-on-SiO$_2$ due to their atomically flat and dangling-bond free surface, and screening of the charge inhomogeneities on the underlying SiO$_2$ \cite{huertas-hernando_spin-orbit-mediated_2009, ertler_electron_2009}. This improvement can be helpful in possibly compensating for the reduced $\tau_{\text{s}}$ due to the enhanced SOC/ spin-absorption \cite{yan_two-dimensional_2016, dankert_electrical_2017}, and improve the spin-signal magnitude, ii) due to partial encapsulation of graphene with the TMD \cite{ghiasi_large_2017,yan_two-dimensional_2016,benitez_strongly_2017}, the encapsulated and non-encapsulated graphene regions have different charge and spin-transport properties. It requires a complex analysis for the accurate interpretation of the TMD induced spin-relaxation in graphene. On the other hand, spin-transport measurements in graphene fully supported on a TMD substrate do not have this drawback and can distinguish the possible effects of spin-absorption via the TMD or a proximity-induced SOC, due to a uniform carrier density and identical effect of the substrate present everywhere in graphene, and iii) in contrast with the TMD-on-graphene geometry \cite{ghiasi_large_2017,yan_two-dimensional_2016,benitez_strongly_2017,dankert_electrical_2017} where graphene partially shields the back-gate induced electric field to the TMD, and one cannot clearly comment on the TMD's conducting state and correlate its effect on spin-transport in graphene, the inverted Gr-on-TMD geometry does not have this drawback. Lastly, it is worth exploring the possibility of recently observed spin-relaxation anisotropy  for in-plane and out-of-plane spins in Gr-TMD heterostructures  \cite{cummings_giant_2017,ghiasi_large_2017,benitez_strongly_2017} in our system.     

\begin{figure*}
\includegraphics[]{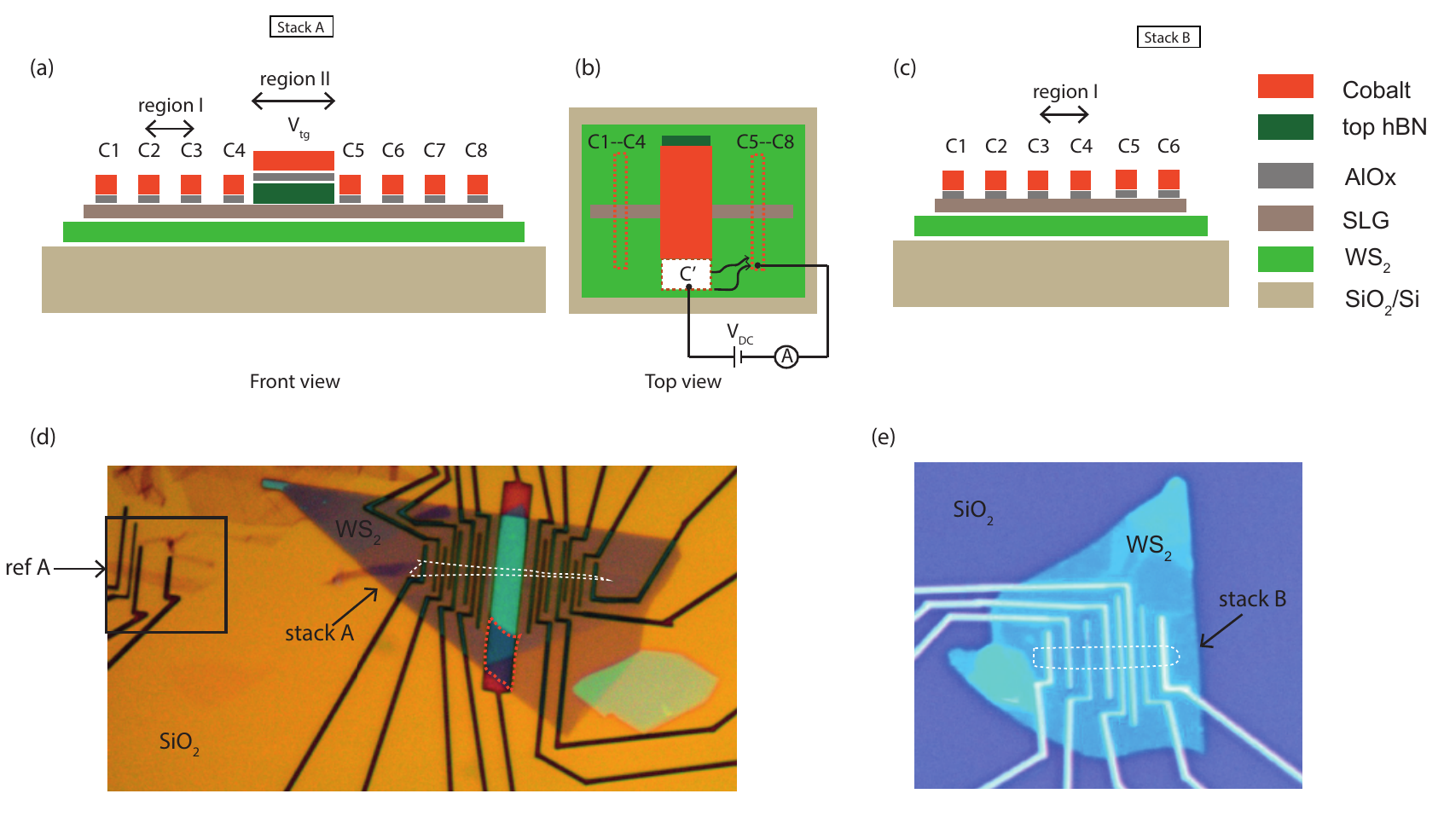}
\caption{\label{geometry} (a) Stack A: a hBN/Gr/WS$_2$ stack with Co/AlO$_{\text{x}}$ ferromagnetic (FM) tunnel contacts and a top gate. (b) Top-view of stack A. White region marked by C' represents the top-gate electrode contacting the WS$_2$ substrate. The connection scheme for measuring the $I-V$ behavior of WS$_2$ is also shown. (c) Stack B: graphene supported on  a bottom WS$_2$ substrate. (d) Optical image of stack A before the contact deposition. The graphene flake is outlined by a white dotted line, and the orange dotted line denotes the WS$_2$ flake region to be contacted by the top-gate electrode after the contact deposition. On the top left corner outlined with a black square, a graphene flake (ref A) with the developed contacts can be seen on the same SiO$_2$/Si substrate. (e) Optical image of stack B, i.e., a  graphene (white dashed lines)/WS$_2$ heterostructure after the contact deposition. It also has a reference Gr flake \textquoteleft ref B' on the same SiO$_2$ substrate (not shown in the image). }
\end{figure*}

We study the charge and spin-transport properties of graphene, fully supported on  a tungsten disulfide (WS$_2$) substrate, and partially encapsulated with a top hBN flake, using a four-probe local and non-local geometry, respectively. We measure large values of charge mobility up to 16,000 cm$^{2}$V$^{-1}$s$^{-1}$. For spin-valve measurements, the obtained spin-signal $\Delta R_{\text{NL}}$ is almost constant and independent of the carrier type and carrier density in graphene, ruling out the possibility of spin-absorption via the underlying WS$_2$ substrate. For Hanle measurements, we obtain a very low spin-relaxation time $\tau_{\text{s}}\sim$ 10 ps in the electron-doped regime than $\tau_{\text{s}}\sim$ 800 ps of a reference graphene flake  on the SiO$_2$/Si substrate in the same chip. Via the top-gate voltage application, we can access the hole doped regime of graphene in the encapsulated region where $\tau_{\text{s}}$ is enhanced up to 40-80 ps for various carrier densities and electric fields. By changing the electric-field in the range of 1 V/nm in the encapsulated region, we can change $\tau_{\text{s}}$ from 20-80 ps, almost by factor of four, which suggests an electric-field controlled Rashba SOC in our system \cite{gmitra_graphene_2015,gmitra_trivial_2016}. For both electron and hole regimes (stronger for the hole regime), we observe the fingerprints of the D'yakonov-Perel' type mechanism for spin-relaxation, similar to WAL measurements \cite{yang_strong_2017, yang_tunable_2016}. For Gr-on-WS$_2$, the ratio of the out-of-plane to the in-plane $\Delta R_{\text{NL}}$ ( therefore $\tau_{\text{s}}$) in the electron-doped regime is less than one, an indicative of an in-plane Rashba-type system \cite{cummings_giant_2017, guimaraes_controlling_2014}. For the hole doped regime, we observe an enhanced out-of-plane spin-signal \cite{benitez_strongly_2017} which suggests a higher $\tau_{\text{s}}^{\perp}$ for the out-of-plane spins. However, in the presence of a similar background magnetoresistance signal, the anisotropic behavior can not be uniquely determined and requires further measurements \cite{ghiasi_large_2017,benitez_strongly_2017}.

We also confirm the signature of WS$_2$ induced SOC in graphene-on-WS$_2$ by measuring the WAL signature, similar to the studies performed in refs.~\cite{wang_origin_2016,wang_strong_2015,yang_strong_2017,yang_tunable_2016}. Therefore, a low $\tau_{\text{s}}$ in graphene-on-WS$_2$ substrate, with an electric-field tunable Rashba SOC and a WAL signature in the same sample can be attributed to the WS$_2$ induced proximity SOC at the graphene-WS$_2$ interface.

\section{Device fabrication}
The graphene-WS$_2$ stacks are prepared on a n$^{++}$-doped SiO$_2$/Si substrate ($t_{\text{SiO}_2}\sim$300 nm) via a dry pick-up transfer method \cite{zomer_fast_2014,omar_graphene-$mathrmws_2$_2017}. The WS$_2$ flake is exfoliated on a polydimethylsiloxane (PDMS) stamp and identified using an optical microscope. The desired flake is transferred onto a pre-cleaned SiO$_2$/Si substrate ($t_{\text{SiO$_2$}}$=300 nm), using a transfer-stage. The transferred flake on SiO$_2$ is annealed in an Ar-H$_2$ environment at 250$^{\circ}$C for 3 hours in order to achieve a clean top-WS$_2$ surface. The graphene (Gr) flake is exfoliated from a ZYB grade HOPG (Highly oriented pyrolytic graphite) crystal and boron nitride (BN) is exfoliated from BN crystals (size$\sim$ 1 mm) onto different SiO$_2$/Si substrates ($t_{\text{SiO$_2$}}$=300 nm). Both crystals were obtained from HQ Graphene. The desired single layer graphene  and hBN flakes are identified using the optical microscope. In order to prepare an hBN/Gr/WS$_2$ stack, first the hBN flake is picked up by a polycarbonate (PC) film attached to a PDMS stamp, using the same transfer-stage.  Next, the Gr flake is aligned with respect to the hBN flake. When graphene is brought in a contact with the hBN flake, the graphene region underneath the hBN flake is picked up by the van der Waals force between the two flakes. The graphene region outside the hBN flake is picked up by the sticky PC film. Now the WS$_2$ flake, previously transferred onto a SiO$_2$/Si substrate, is aligned and brought in a contact with the PC/hBN/Gr assembly and the whole system is heated up to 150$^{\circ}$C, so that the PC/hBN/Gr assembly is released onto the WS$_2$ substrate. Now, the stack is put in a chloroform solution for 3 hours in order to remove the  PC film used in the stack preparation. After that, the stack is annealed again in the Ar-H$_2$ environment for five hours at 250$^{\circ}$C to remove the remaining polymer residues. The thicknesses of WS$_2$ and BN flakes were characterized by the Atomic Force Microscopy measurements.

In order to define the contacts, a poly-methyl methacrylate (PMMA) solution is spin-coated over the stack and the contacts are defined via the electron-beam lithography (EBL). The PMMA polymer exposed via the electron beam gets dissolved in a MIBK:IPA (1:3) solution. In the next step, 0.7 nm Al is deposited in two steps, each step of 0.35 nm followed by 12 minutes oxidation in an oxygen environment to form a AlO$_x$ tunnel barrier. On top of it, 70 nm thick cobalt (Co) is deposited to form the ferromagnetic (FM) tunnel contacts with a 3 nm thick Al capping layer to prevent the oxidation of Co electrodes, followed by the lift-off process in acetone solution at 30$^{\circ}$C.


\section{Results}
We study two samples: i) stack A: a  hBN/Gr/WS$_2$ stack consisting of a single layer graphene  encapsulated between a bottom-WS$_2$ ($t_{\text{WS$_2$}}\sim$ 3.5 nm) and a top-hBN ($t_{\text{hBN}} \sim$ 8 nm) flake , as shown in Figs.~\ref{geometry}(a,b,d)  and ii) stack B: a WS$_2$/Gr stack consisting of a single layer graphene supported on a bottom WS$_2$ flake ($t_{\text{WS$_2$}}\sim$ 4.2 nm), without any hBN encapsulation from the top, as shown in Figs.~\ref{geometry}(c,e).
 On the same SiO$_2$/Si chip, there are reference graphene flakes near stack A (Fig.~\ref{geometry}(d)) and stack B. Therefore, we can directly compare the charge and spin-transport properties of the reference Gr flakes on SiO$_2$ and graphene-on-WS$_2$ substrate, prepared via identical steps. The reference flakes on the same SiO$_2$, shared by stack A and stack B, are labeled as \textquoteleft ref A' and \textquoteleft ref B' respectively. Moreover, stack A  has non-encapsulated regions (region-I) and an encapsulated region (region-II) both, as indicated in the device schematic of Fig.~\ref{geometry}(a). On the other hand, stack B only consists of region-I. Therefore, we will discuss the data of stack A as a representative device. 

 We use a low-frequency lock-in detection method to measure the charge and spin transport properties of the graphene flake. In order to measure the I-V behavior of the WS$_2$ flake and for gate-voltage application, a Kiethley 2410 dc source meter was used.
All measurements are performed at room temperature and at 4 K under the vacuum conditions in a cryostat.
\subsection{Charge transport measurements}  

We measure the charge transport via the four-probe local measurement scheme. For measuring the gate-dependent resistance, i.e., the Dirac behavior of graphene-on-WS$_2$ in region-I (II) of stack A, a fixed  ac current $i_{\text{ac}} \sim$ 100 nA is applied between contacts C1-C4 (C1-C6) and the voltage-drop is measured between contacts C2-C3 (C4-C5), while the back-gate (top-gate) voltage is swept. The maximum resistance point in the Dirac curve is denoted as the charge neutrality point (CNP). For graphene-on-WS$_2$, it is possible to tune the Fermi energy $E_{\text{F}}$ in graphene  until $E_{\text{F}}$ lies within the band-gap of WS$_2$. After $E_{\text{F}}$ coincides with the conduction band edge of WS$_2$, it also starts conducting, and $V_{\text{bg}}$ corresponding to this transition is denoted as $V_{\text{on}}$. For $V_{\text{bg}}>V_{\text{on}}$, the WS$_2$ flake screens the electric field from the back-gate due to a charge accumulation at the SiO$_2$-WS$_2$ interface \cite{wang_strong_2015} and the resistance of the graphene flake cannot be further modified via $V_{\text{bg}}$. 
\begin{figure*}
\includegraphics[]{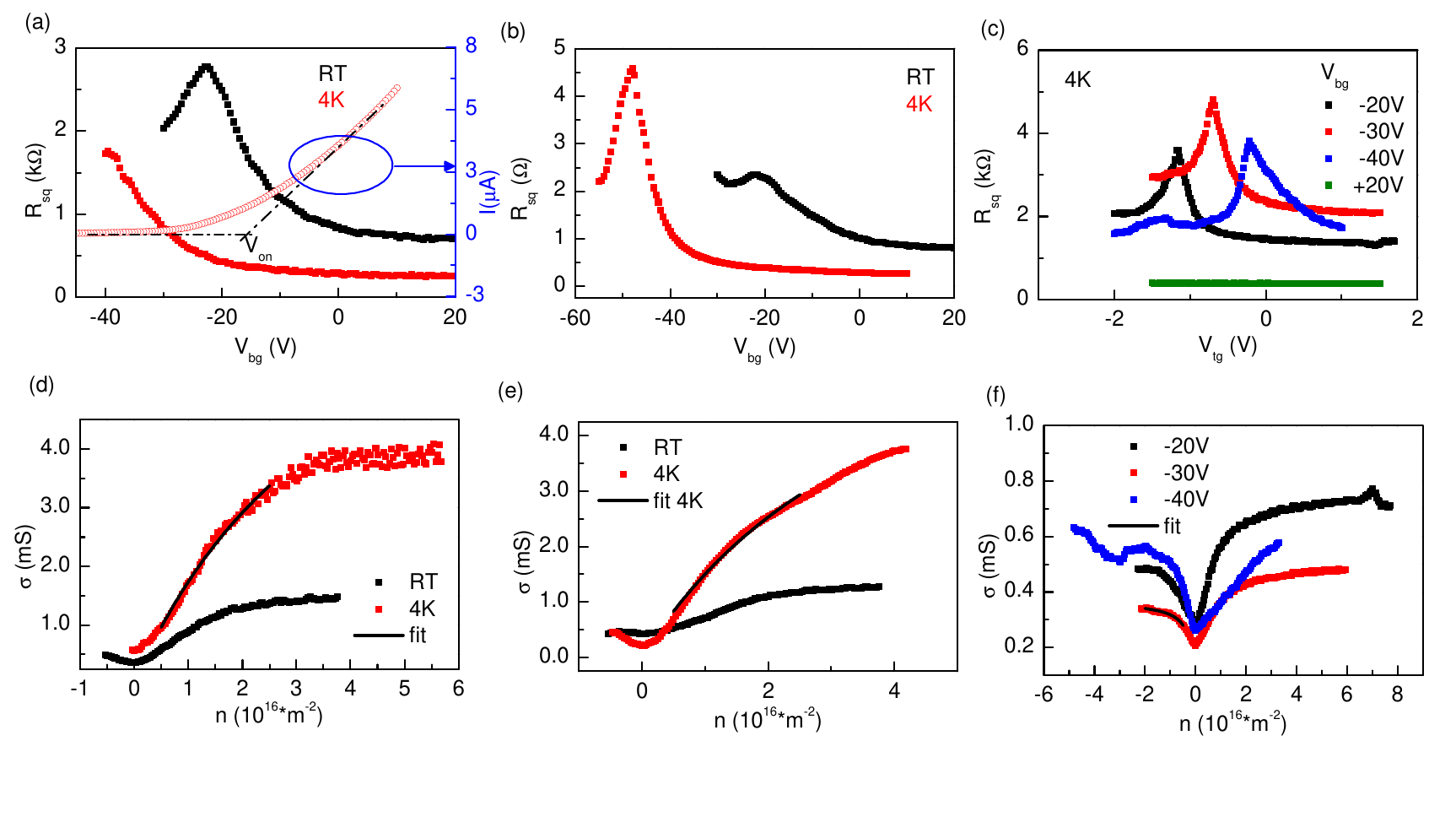}
\caption{\label{fig1} (a) For region-I of stack A, the $R_{\text{sq}}-V_{\text{bg}}$ dependence at RT and 4 K is shown on the left axis. The $I_{\text{DS}}-V_{\text{bg}}$ behavior of WS$_2$ at 4 K is shown on the right-y axis (open circle).  For region-II (b) the $R_{\text{sq}}-V_{\text{bg}}$ and (c) the $R_{\text{sq}}-V_{\text{tg}}$ behavior of Gr encapsulated between WS$_2$ and hBN flakes. The corresponding $\sigma-V_{\text{bg(tg)}}$ behaviors are plotted in (d), (e) and (f).}
\end{figure*}

The Dirac curves for region-I and region-II of stack A are shown as a function of $V_{\text{bg}}$ in Fig~\ref{fig1}(a) and Fig~\ref{fig1}(b), respectively.  The same is also shown as a function of top-gate voltage $V_{\text{tg}}$ in region-II in Fig.~\ref{fig1}(c). In order to extract the carrier mobility $\mu$, we fit the charge-conductivity $\sigma$ versus carrier density $n$ plot with the following equation:
\begin{equation}
 \sigma=\frac{1}{R_{\text{sq}}}=\frac{ne\mu+\sigma_0}{1+R_{\text{s}}(ne\mu+\sigma_0)}.
 \label{mobility}
\end{equation}
 Here $R_{\text{sq}}$ is the square resistance of graphene, $\sigma_0$ is the conductivity at the CNP, $R_{\text{s}}$ is the residual resistance due to short-range scattering \cite{morozov_giant_2008, gurram_spin_2016,zomer_fast_2014} and $e$ is the electronic charge. We fit the $\sigma-n$ data for $n$ (both electrons and holes) in the range 0.5-2.5$\times$10$^{12}$ cm$^{-2}$ with Eq.~\ref{mobility}. For the non-encapsulated region we obtain the electron-mobility $\mu_{\text{e}}\sim$ 9,700 cm$^2$V$^{-1}$s$^{-1}$ at room temperature (RT), which is enhanced up to 13,400 cm$^2$V$^{-1}$s$^{-1}$ at 4 K (Fig.~\ref{fig1}(d)). For the encapsulated region, we extract a relatively lower $\mu_{\text{e}} \sim$ 7,300 cm$^2$V$^{-1}$s$^{-1}$ at RT which is enhanced at 4 K up to 11,500 cm$^2$V$^{-1}$s$^{-1}$ (Fig.~\ref{fig1}(e)). Via the top gate voltage application, we can access the hole carrier densities up to $\sim$ -7$\times$10$^{16}$cm$^{-2}$, and extract the hole mobility $\mu_{\text{h}}$ at different $V_{\text{bg}}$ values in the range 12,600-16,000 cm$^2$V$^{-1}$s$^{-1}$ at 4 K (Fig.~\ref{fig1}(f)). Via this analysis, we get  $\mu_{\text{e}}\sim$ 6,000-13,000 cm$^2$V$^{-1}$s$^{-1}$ at different $V_{\text{bg}}$ values, similar to values that were extracted from the back-gate sweep in Fig.~\ref{fig1}(e). 
 
 In order to obtain the transfer characteristics of the WS$_2$ substrate, we use a specific measurement geometry. Due to partial encapsulation of the bottom-WS$_2$ via the top-hBN layer, as marked by the orange dashed lines in Fig.~\ref{geometry}(d), the WS$_2$ crystal is contacted via the top gate electrode (white region in Fig.~\ref{geometry}(b), labeled as C') and one of the electrodes C1--C8 on the graphene flake. For a voltage applied between C' and C$_j$ ($j=1-8$), there is a current flowing through WS$_2$, as schematically indicated by arrows in Fig.~\ref{geometry}(b). The $I_{\text{DS}}-V_{\text{bg}}$ transfer curve for WS$_2$ measured using this geometry is plotted in Fig.~\ref{fig1}(a) (marked by blue ellipse). It is also noteworthy that there is a negligible gating action in graphene from the top gate  when the WS$_2$ is conducting at $V_{\text{bg}}$=+20 V (Fig.~\ref{fig1}(c)).
 
 In conclusion, for graphene-on-WS$_2$, we obtain high electron and hole mobilities reaching up to 16,000 cm$^2$V$^{-1}$s$^{-1}$. We obtain similar mobilities for both encapsulated and the non-encapsulated regions, implying that the observed high mobility is due to a clean Gr-WS$_2$ interface in our samples, and is not significantly affected by the lithographic process during the sample preparation.

\subsection{Spin-transport measurements}

A nonlocal four-probe connection scheme is used to measure the spin-transport in graphene. In order to measure the spin signal $\Delta R_{\text{NL}}$ in the non-encapsulated(encapsulated) region, $i_{\text{ac}}$ is applied between contacts C2-C1(C4-C1) and the nonlocal voltage $v_{\text{NL}}$ is measured between C3-C8(C5-C8),  in Fig.~\ref{geometry}(a) \cite{tombros_electronic_2007}. 

\begin{figure*}
\includegraphics[]{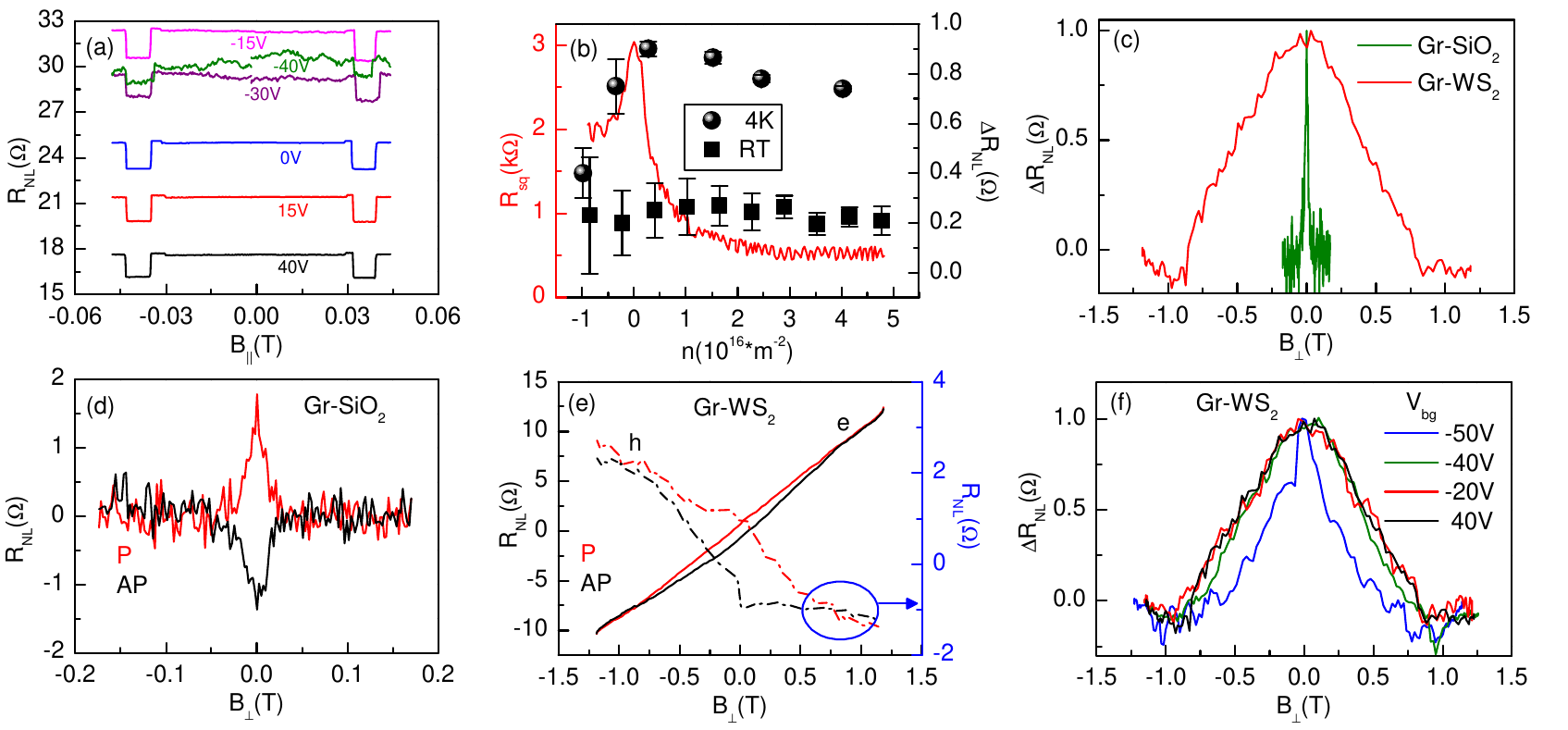}
\caption{\label{fig2}(a) Spin-valve measurements for Gr-on-WS$_2$ (region-I of stack A) at different $V_{\text{bg}}$ for the injector-detector separation $L$=0.8 $\mu$m, and the corresponding (b) $\Delta R_{\text{NL}}$ as a function of carrier density in graphene at RT and 4 K. (c) Normalized Hanle signal $\Delta R_{\text{NL}}(B_{\perp})$ for graphene-on-SiO$_2$ (green) and on-WS$_2$ (red) at 4 K. (d) Parallel (P) and anti-parallel (AP) Hanle signals $R_{\text{NL}}$ for graphene-on-SiO$_2$ and (e) for graphene-on-WS$_2$ (region-I of stack A). A large linear background  can also be seen in both P and AP configurations and in electron and hole-doped regimes. (f) $\Delta R_{\text{NL}}(B_{\perp})$ in region-I of stack A at different $V_{\text{bg}}$ at 4 K.}
\end{figure*}

For spin-valve measurements, first an in-plane magnetic field $B_{||}\sim$ 0.2 T is applied along the easy axes of the ferromagnetic (FM) electrodes, so that they have their magnetization aligned in the same direction. The FM contacts are designed with different widths, therefore they have different coercivities. Now,  $B_{||}$ is swept in the opposite direction, and depending on their coercivities, the FM contacts reverse their magnetization direction along the applied field, one at a time. This magnetization reversal appears as a sharp transition in $v_{\text{NL}}$ or in the nonlocal resistance $R_{\text{NL}}=v_{\text{NL}}/i_{\text{ac}}$
, as shown in Figs.~\ref{fig2}(a) and \ref{fig3}(a). The spin-signal is $\Delta R_{\text{NL}}=\frac{R_{\text{NL}}^P-R_{\text{NL}}^{AP}}{2}$, where $R_{\text{NL}}^{P(AP)}$ represents the $R_{\text{NL}}$ value of the two level spin-valve signal, corresponding to the parallel (P) and anti-parallel (AP) magnetization of the FM electrodes.

For Hanle spin-precession measurements, first the FM electrodes are magnetized in the parallel (P) or anti-parallel (AP) configuration. Next, for a fixed P (AP) configuration, an out-of-plane magnetic field $B_{\perp}$ is applied and the injected spin-accumulation precesses around the applied field with the Larmor frequency $\overrightarrow{\omega_{\text{L}}}= \frac{g \mu_{\text{B}}}{\hbar}{B_{\perp}}$,  while diffusing towards the detector, and gets dephased. Here $g$ is the gyromagnetic ratio(=2) for an electron, $\mu_{\text{B}}$ is the Bohr magneton and $\hbar$ is the reduced Planck constant.  The measured Hanle curves are fitted with the steady state solution to the one-dimensional Bloch equation \cite{tombros_electronic_2007}:
\begin{equation}
 D_{\text{s}} {\bigtriangledown}^2\overrightarrow{\mu_{\text{s}}}-\frac{\overrightarrow{\mu_{\text{s}}}}{\tau_{\text{s}}}+\overrightarrow{\omega_{\text{L}}}\times \overrightarrow{\mu_{\text{s}}}=0
\label{bloch}
\end{equation}
 with the spin diffusion constant $D_{\text{s}}$, spin relaxation time $\tau_{\text{s}}$ and spin-accumulation $\overrightarrow{\mu_{\text{s}}}$ in the transport channel. The spin diffusion length $\lambda_{\text{s}}$ is $ \sqrt{D_{\text{s}}\tau_{\text{s}}}$. 
 
 Hanle measurements for ref A sample are shown in Fig.~\ref{fig2}(d). Since we do not observe the CNP, we could only measure the spin-transport only in the electron-doped regime and obtain $D_{\text{s}}\sim$ 0.02 m$^2$s$^{-1}$ and $\tau_{\text{s}}$ in the range 730-870 ps, i.e., $\lambda_{\text{s}} \sim$ 3.6-3.8 $\mu$m.

\begin{figure}
  \includegraphics[]{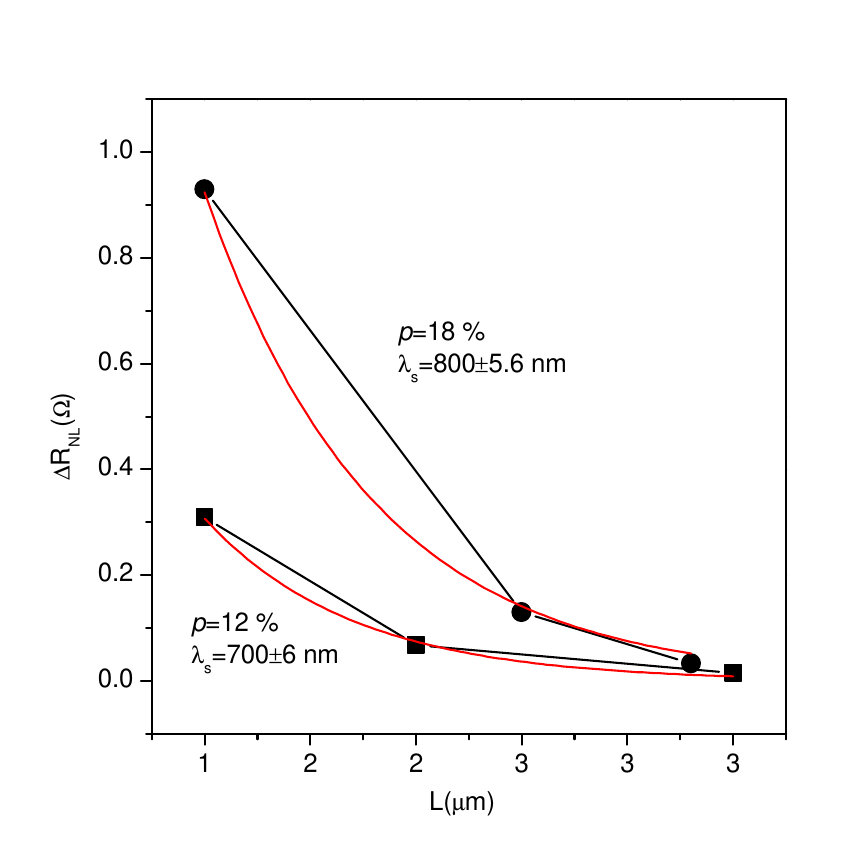}
  \caption{\label{nl distance} Exponentially decaying spin signal $\Delta R_{\text{NL}}$ in stack A (region-I) for an increasing injector-detector separation $L$. Black square and circle data points are taken for two different injector electrodes. Here, we assume equal spin-polarization for all the contacts. The data is fitted using Eq.~\ref{rnl}.}
\end{figure} 
 After obtaining the spin-transport parameters for ref A, we measure the spin-transport in graphene-on-WS$_2$ substrate (region I of stack A) on the same chip. For a varying range of carrier density in graphene, from electron to hole regime with the application of $V_{\text{bg}}$, we measure almost a constant spin signal $\Delta R_{\text{NL}}$ at RT via spin-valve measurements, plotted in Fig.~\ref{fig2}(b). At 4 K, the spin signal shows a modest increase around the CNP, and then it decreases. For $V_{\text{bg}}<$ -30 V, there is a negligible in-plane charge conduction in WS$_2$ (Fig.~\ref{fig1}(a)). If the spin-absorption via WS$_2$ was the dominant spin-relaxation mechanism, the spin-signal should enhance for $V_{\text{bg}}<$ -30 V. Both observations cannot be explained by considering the gate-tunable spin-absorption as a dominant source of spin-relaxation at the graphene-WS$_2$ interface within the applied $V_{\text{bg}}$ range. 
 
 Now we perform spin-valve measurements in the encapsulated region (region-II of stack A), as a function of  $V_{\text{bg}}$ and $V_{\text{tg}}$ (Fig.~\ref{fig3}(a)). For a wide range of carrier density in the encapsulated graphene which is equivalent to applying $V_{\text{bg}}$ in the range of $\pm$60 V, we do not see any significant change in the spin-signal in Fig.~\ref{fig3}(a), similar to the back-gate dependent spin-valve measurements (Fig.~\ref{fig2}(a)). It  leads to a conclusion that $\Delta R_{\text{NL}}$ is independent of the carrier density, carrier type in graphene and the conducting state of the TMD. Note that this configuration is similar to the TMD-on-graphene configuration with a back gate application in ref.\cite{yan_two-dimensional_2016,dankert_electrical_2017}, except graphene is uniformly covered with the WS$_2$ flake in our sample.
 
 In order to estimate $\lambda_{\text{s}}$ from spin-valve measurements in region-I, we measure $\Delta R_{\text{NL}}$ at different injector-detector separation $L$ values.  Assuming equal polarization $p$ for all the contacts, we can estimate $\lambda_{\text{s}}$ using the relation \cite{tombros_electronic_2007}:
 \begin{equation}
  \Delta R_{\text{NL}}=\frac{p^2R_{\text{sq}}\lambda_{\text{s}}e^{-\frac{L}{\lambda_{\text{s}}}}}{2w}
  \label{rnl}
 \end{equation}
where $w$ is the width of the spin-transport channel.  We obtain $\lambda_{\text{s}}$ around 700-800 nm (Fig.~\ref{nl distance}), which is almost five times lower than $\lambda_{\text{s}}$ in ref A sample. For graphene-on-WS$_2$, we obtain the charge diffusion coefficient $D_{\text{c}}\sim$ 0.05 m$^2$s$^{-1}$  using the Einstein relation: $\sigma=e^2D_{\text{c}}\nu$, where $\nu$ is the density-of-states in graphene.  Assuming $D_{\text{s}}=D_{\text{c}}$ \cite{guimaraes_controlling_2014}, we estimate $\tau_{\text{s}} \sim$ 10 ps, using $\lambda_{\text{s}}$ obtained from spin-valve measurements (Fig.~\ref{nl distance}).  Note that this value may be uncertain due to different polarization values of the individual contacts, still it gives an estimate of $\lambda_{\text{s}}$ \cite{popinciuc_electronic_2009}.  

In region-I of stack A, we measure broad Hanle curves with full-width half maximum in the range of $\sim$  1 T (Figs.~\ref{fig2}(c),(e), (f)). A direct comparison between Hanle curves of the reference sample and for graphene-on-WS$_2$, plotted together in  Fig.~\ref{fig2}(c), clearly demonstrates the effect of the WS$_2$ substrate in the broadening of the Hanle curve. The line shape of $\Delta R_{\text{NL}}$ remains similar at different carrier densities ($n \sim$0-6$\times$10$^{16}$ m$^{-2}$) in the electron-doped regime (Fig.~\ref{fig2}(f)). Note that the WS$_2$ gets switched on around the CNP of the graphene and remains in the conducting-state in this regime. By using the Hanle curve fitting procedure, we obtain $\tau_{\text{s}} \sim$ 10-13 ps and $D_{\text{s}} \sim$ 0.03-0.04 m$^2$s$^{-1}$ which matches with $D_{\text{c}}$ within factor of two obtained from the charge transport measurements. With the obtained $D_{\text{s}}$ and $\tau_{\text{s}}$ via the Hanle measurements, we achieve $\lambda_{\text{s}} \sim$ 600-700 nm, using $\lambda_{\text{s}}=\sqrt{D_{\text{s}}\tau_{\text{s}}}$, in a good-agreement with $\lambda_{\text{s}}$ obtained from the distance-dependence of spin-valve measurements. In the hole transport regime, we could perform the Hanle measurements only at $V_{\text{bg}}$~=~-50V ({$D_{\text{s}}\sim$ 0.35 m$^2$s$^{-1}$, $\tau_{\text{s}} \sim$ 35 ps) with $D_{\text{s}}$ and $D_{\text{c}}$ ($\sim$0.03 m$^2$s$^{-1}$) differing by an order of magnitude. Therefore, we cannot comment on the spin-transport parameters in the hole transport regime in region-I.
   \begin{figure*}
\includegraphics[]{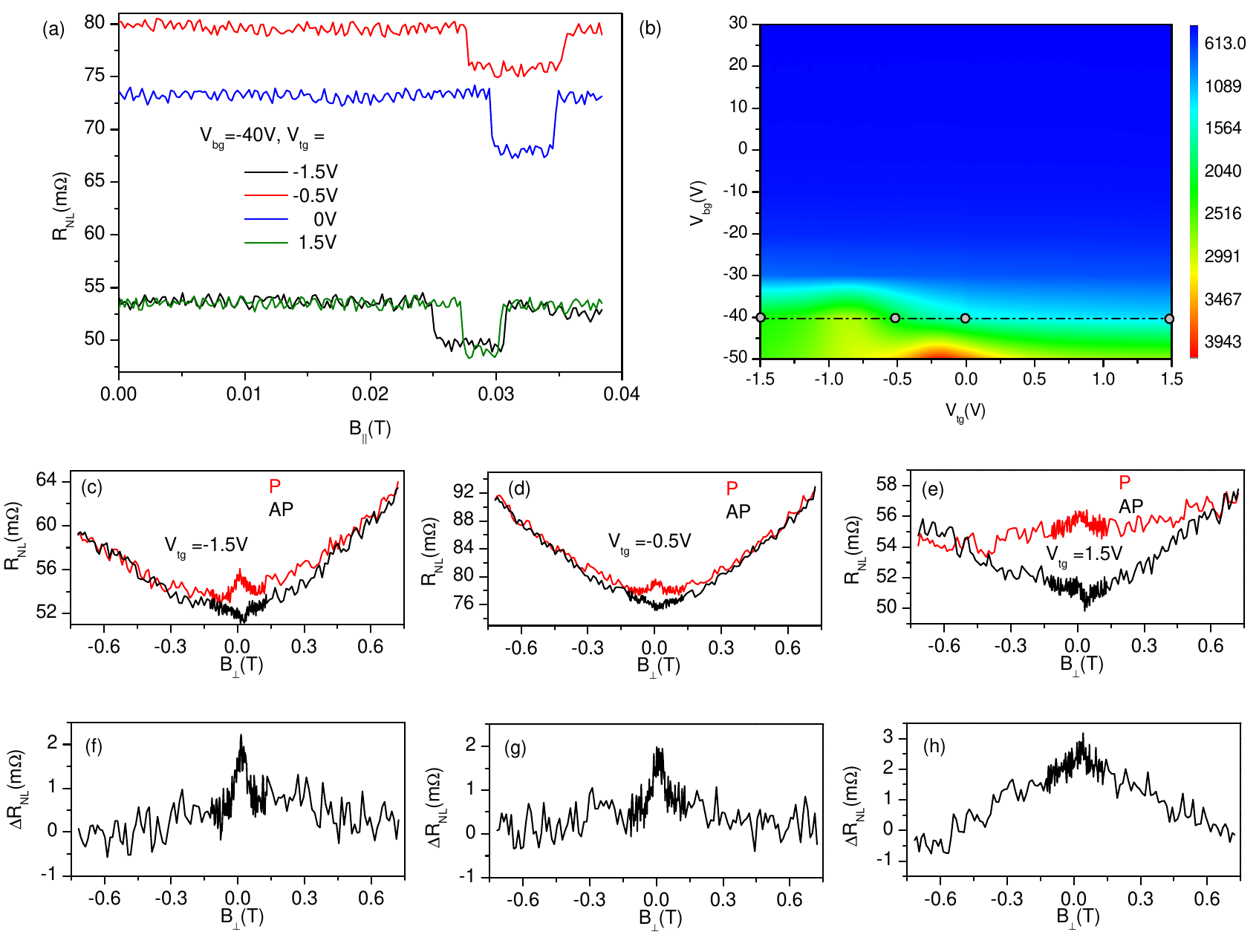}
\caption{\label{fig3}(a) Spin-valve measurements across the encapsulated region (region-II) of stack A at different top-gate voltages, changing the carrier-density of the encapsulated region from hole to electron-doped regime. (b) A contour-plot of $R_{\text{sq}}$ for the encapsulated region as a function of $V_{\text{bg}}$ and $V_{\text{tg}}$. The gray circles on the horizontal dotted line at $V_{\text{bg}}$=-40 V denote the $V_{\text{tg}}$ values at which spin valve and Hanle measurements are taken.  Hanle measurements for the encapsulated region for the hole doped regime, at the CNP and electron-doped regime are shown in (c), (d) and (e), respectively. The corresponding Hanle signals are shown in (f), (g) and (h).} 
\end{figure*}

It should be noted that at high  out-of-plane magnetic fields $B_{\perp}\sim$1 T, the magnetization direction of the FM electrodes does not fully lie in the sample-plane and a makes an angle with the plane \cite{guimaraes_controlling_2014}. When we correct the measured data for the angular ($B_{\perp}$) dependence of the magnetization (not shown here) using the procedure in ref.\cite{isasa_origin_2016}, the \textquoteleft corrected' Hanle curves become even broader. From these Hanle curves, we would obtain even lower $\tau_{\text{s}}$. Therefore, the $\tau_{\text{s}}$ values reported here represent the upper bound. 

We estimate the contact polarization $p$  $\sim$ 15-20 \% using Eq.~\ref{rnl} for this device which along with a reasonably good $D_{\text{s}} \sim$ 0.04 m$^2$s$^{-1}$, enables us to measure a large $\Delta R_{\text{NL}}$ in the order of Ohms, even with such a short $\tau_{\text{s}}$. For stack B, we obtain a small $p\sim$ 1-3\% and therefore a small  $\Delta R_{\text{NL}} \sim$ 7 m$\Omega$, making it difficult to measure clear Hanle signals at high magnetic fields in the presence of a large linear background. 

For individual Hanle curves measured in P or AP configuration, we also observe a large linear background signal ($\sim$ 10-20 $\Omega$) along with the Hanle signal (Fig.~\ref{fig2}(e)). The sign of the background-slope changes with respect to the change in the carrier-type from electrons to holes, similar to a Hall-like signal \cite{volmer_contact-induced_2015}. However, we do not expect such a large Hall background because the FM electrodes are designed across the width of the graphene flake. The source of such background is non-trivial and at the moment is not clear to us.

\section{Discussion} 
\begin{figure*}
 \includegraphics[]{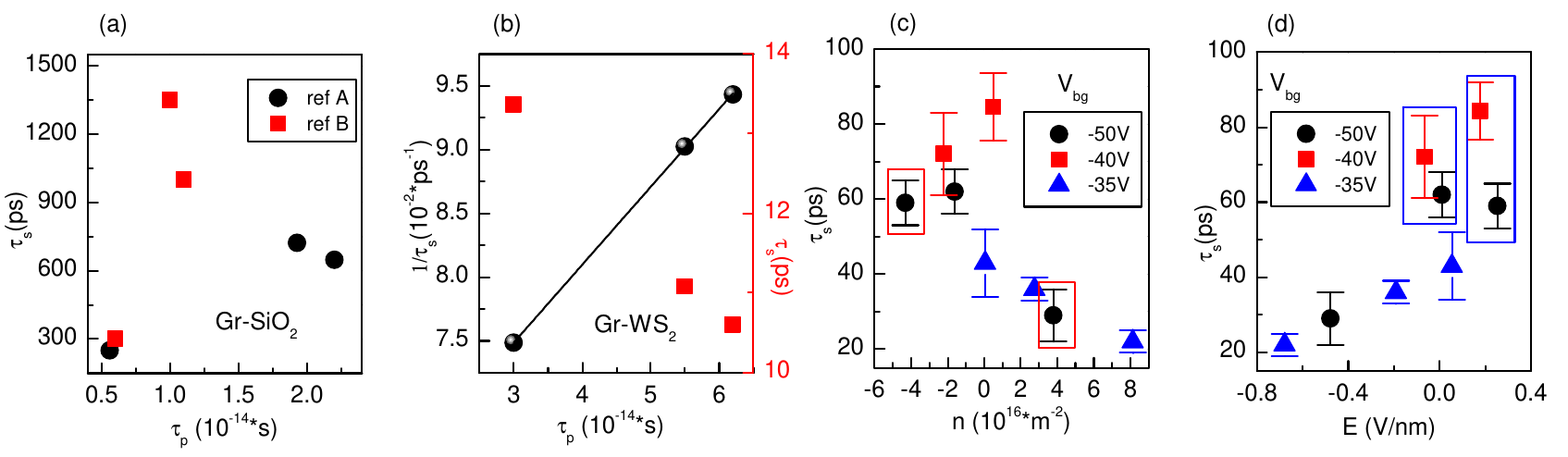}
 \caption{\label{taus}(a) $\tau_{\text{s}}$ versus $\tau_{\text{p}}$ for the reference graphene on SiO$_2$ substrate in the electron doped-regime shows an enhanced $\tau_{\text{s}}$ with the increase in $\tau_{\text{p}}$, suggesting the EY-type spin-relaxation. (b) $\tau_{\text{s}}$ versus $\tau_{\text{p}}$ (red squares) for graphene-on-WS$_2$ substrate (region-I of stack A) shows an enhanced $\tau_{\text{s}}$ for a reduced $\tau_{\text{p}}$, suggesting the DP-type spin relaxation in presence of a substrate induced SOC. Black line represents a linear fit of $1/\tau_{\text{s}}-\tau_{\text{p}}$ data (black spheres). (c) $\tau_{\text{s}}$ as a function of carrier density $n$ and (d) Electric field $E$ at different values of $V_{\text{bg}}$ for the electron and hole transport regime in region-II of stack A. $E$ and $n$ in the encapsulated-region due to a combined effect of the top and bottom gates are calculated by following the procedure in ref.~\cite{guimaraes_controlling_2014}.}
\end{figure*}

 In graphene, there are two dominant spin-relaxation mechanisms \cite{han_graphene_2014, han_spin_2011, jozsa_linear_2009} : 1) Elliot-Yafet (EY) mechanism where an electron-spin is scattered during the interaction with the impurities. Therefore, the spin-relaxation time is proportional to the momentum relaxation time $\tau_{\text{p}}$, i.e., $\tau_{\text{s}} \propto \tau_{\text{p}}$, 2) D'yakonov-Perel' (DP) mechanism, where the electron-spin precesses in a spin-orbit field between two momentum scattering events,  following the relation $\tau_{\text{s}}\propto \frac{1}{\tau_{\text{p}}}$.  

 In order to check the relative contribution of the EY and DP mechanisms in our samples,  we plot the $\tau_{\text{s}}$ versus $\tau_{\text{p}}$ dependence. Here, $\tau_{\text{p}}$  is calculated from the diffusion coefficient $D$, using the relation $D\sim {v_{{\text{F}}}^2\tau_{\text{p}}}$, assuming $D=D_{\text{s}}(D_{\text{c}})$.  For reference samples on the SiO$_2$ substrate, $\tau_{\text{s}}$ increases with $\tau_{\text{p}}$ in the electron doped regime (Fig.~\ref{taus}(a)), suggesting the dominance of the EY-type spin relaxation in single layer graphene on the SiO$_2$ substrate, similar to previous observations \cite{jozsa_linear_2009, popinciuc_electronic_2009, han_spin_2011} on this system. We could not quantify the spin-orbit strength due to unknown carrier density and the corresponding Fermi energy \cite{zomer_long-distance_2012}. For stack A (region-I), processed in identical conditions, we observe an opposite trend between $\tau_{\text{s}}$ and $\tau_{\text{p}}$ in the electron-doped regime (Fig.~\ref{taus}(b)), which resembles the DP type mechanism. We fit the data with the relation $\frac{1}{\tau_{\text{s}}}=\frac{4\lambda_{\text{R}}^2}{\hbar^2}\tau_{\text{p}}$ \cite{yang_tunable_2016} and extract the Rashba SOC strength $\lambda_{\text{R}}\sim$ 250 $\mu$eV, which is 4 to 6 times higher than the spin orbit coupling strength in a similar mobility graphene-on-hBN substrate reported in ref.~\cite{zomer_long-distance_2012}, and distinguishes the effect of WS$_2$ substrate in enhancing the SOC in graphene. The obtained magnitude of $\lambda_{\text{R}}$ is of similar order as reported in refs.~\cite{cummings_giant_2017, yang_tunable_2016, wang_origin_2016, gmitra_graphene_2015}.  However, a slight variation in $\tau_{\text{s}}$ can drastically change the $\tau_{\text{s}}-\tau_{\text{p}}$ dependence and thus the value of $\lambda_{\text{R}}$. Therefore, such a small variation of the spin-relaxation rate ($\tau_{\text{s}}^{-1}$) from 75 ns$^{-1}$ to 95 ns$^{-1}$ restricts us from claiming the dominance of the DP spin-relaxation via this analysis. 

 Now we perform Hanle spin-precession measurements in the encapsulated graphene ($L \sim$ 6.7 $\mu$m region-II of stack A). Due to the partial encapsulation of WS$_2$ via hBN (Fig.~\ref{geometry}(b),(d)), we can study the effect of the top-gate on the spin-transport  only when the bottom-WS$_2$ does not conduct. For a fixed $V_{\text{bg}}<$ -30V, we can access both  electron and hole regimes via the top gating. Hanle measurements shown in Figs.~\ref{fig3}(c)-(h) at $V_{\text{bg}}$=-40 V correspond to the CNP of the back-gated graphene, while varying $V_{\text{tg}}$ from the hole-doped regime at $V_{\text{tg}}$=-1.5 V to the electron-doped regime at $V_{\text{tg}}$=+1.5 V.  Here, we can control the carrier-density and electric field in the encapsulated region. An out-of-plane electric field breaks the z$\leftrightarrow$-z inversion symmetry in graphene and modifies the in-plane Rashba SOC \cite{min_intrinsic_2006, gmitra_band-structure_2009, guimaraes_controlling_2014}. For the hole regime at $V_{\text{tg}}$=-1.5 V (Figs.~\ref{fig3}(c),(f)) and -0.5 V (Figs.~\ref{fig3}(d),(g)), we measure a narrow Hanle shape, an indicative of a higher $\tau_{\text{s}}$ for the hole spins.  Now, at $V_{\text{tg}}$=+1.5 V, when the encapsulated region is in the electron-doped regime, the broad Hanle corresponding to a lower $\tau_{\text{s}}$ appears (Figs.~\ref{fig3}(e),(h)). This feature is consistent with broad Hanle curves measured in the electron-doped regime of region-I (Figs.~\ref{fig2}(c),(e),(f)). We fit the Hanle data in the hole-doped regime for $|B_{\perp}|<$ 200 mT, while assuming $D_{\text{s}}=D_{\text{c}}$, where $D_{\text{c}}$ is obtained from the $\sigma-V_{\text{tg}}$ dependence in Figs.~\ref{fig1}(f), and obtain $\tau_{\text{s}} \sim$ 40-80 ps. We repeat the Hanle measurements for $V_{\text{bg}}$=-50 V,-35 V and observe a similar behavior, confirming that the hole and electron spins have different $\tau_{\text{s}}$ values with $\tau_{\text{s}}^{\text{h}} > \tau_{\text{s}}^{\text{e}}$ (Fig.~\ref{taus}(c)), where superscripts h and e refer to holes and electrons, respectively. By modulating $E_{\perp}$ in the range of 1 V/nm, we can change $\tau_{\text{s}}$ almost by factor of four, which demonstrates the effective control of electric field in changing the SOC, and therefore $\tau_{\text{s}}$ at the Gr-WS$_2$ interface \cite{gmitra_graphene_2015,gmitra_trivial_2016}.
 
 A higher $\tau_{\text{s}}^{\text{h}}$ in the encapsulated region is possibly due to  a combined effect of an intrinsically reduced spin-orbit coupling in the hole regime \cite{cummings_giant_2017, gmitra_trivial_2016} and modification of the electric-field induced Rashba SOC \cite{guimaraes_controlling_2014, yang_strong_2017, yang_tunable_2016}. This can be seen in two features evident from Fig.~\ref{taus}(c) and Fig.~\ref{taus}(d). First, for a similar carrier density magnitude in the electron and hole regime, a reduced $\tau_{\text{s}}$ is observed in the electron-doped regime. Here, the electric field $E$ is pointing towards WS$_2$, i.e., $E<0$ (red box in Fig.~\ref{taus}(c)). Second, for the same electric field an enhanced $\tau_{\text{s}}$ is observed at lower carrier densities (blue box in Fig.~\ref{taus}(d)), similar to that obtained from the WAL experiments in refs.~\cite{yang_strong_2017, wang_origin_2016}. These observations support the presence of a DP type spin-relaxation mechanism for the hole transport and an electric field controllable SOC at the graphene-WS$_2$ interface. 
 
 Recently an anisotropic spin-relaxation, i.e., a higher $\tau_{\text{s}}$ for the spins oriented perpendicular to the graphene plane than in the graphene plane ($\tau_{\text{s}}^{\perp} >\tau_{\text{s}}^{||}$) in graphene-TMD heterostructures was theoretically predicted by Cummings \emph{et al.} \cite{cummings_giant_2017} and was subsequently experimentally demonstrated \cite{ghiasi_large_2017, benitez_strongly_2017}. In order to check this possibility in our system, we subtract the linear background (Figs.~\ref{fig2}(e)) from the measured Hanle data for region-I. In the electron-doped regime, the out-of-plane to in-plane spin signal ratio is always less than one, implying $\tau_{\text{s}}^{\perp} <\tau_{\text{s}}^{||}$. It could be due to the presence of a dominant in-plane Rashba SOC \cite{guimaraes_controlling_2014} in our system. However, in the hole-doped regime of region-II, we observe an increase in $R_{\text{NL}}$ for a high $B_{\perp}$ (Figs.~\ref{fig3}(c),(d)), which is along the lines of a gate-tunable anisotropy in $\tau_{\text{s}}$ reported by Benitez \emph{et al.} \cite{benitez_strongly_2017}. In order to confirm the origin of the enhanced $R_{\text{NL}}$, we measure the magnetoresistance (MR) of the encapsulated region as a function of $B_{\perp}$ and obtain a similar order of change in the graphene-MR. Therefore, we cannot unambiguously determine the presence of an anisotropic spin-relaxation in our system, and additional Hanle measurements as a function of in-plane \cite{ghiasi_large_2017} and oblique magnetic field \cite{benitez_strongly_2017} will be required to draw a conclusion. 

 According to Cummings \emph{et al.} \cite{cummings_giant_2017}, the anisotropy in the in-plane and out-of plane spin-relaxation can not always be observed. It depends on the intervalley scattering rate and the relative strengths of  the in-plane Rashba SOC $\lambda_{\text{R}}$ induced at the graphene-WS$_2$ interface due to broken inversion symmetry \cite{cummings_giant_2017, yang_tunable_2016,yang_strong_2017} and the out-of-plane valley-Zeeman SOC $\lambda_{\text{V}}$ induced in graphene due to the intrinsic SOC in WS$_2$ \cite{cummings_giant_2017, yang_tunable_2016,wang_strong_2015}. In case of a weak-intervalley scattering, the dominant Rashba SOC gives rise to a faster relaxation of the out-of-plane spins and hinders us from observing a strong anisotropic effect \cite{cummings_giant_2017}. However, a direct conclusion regarding the intervalley scattering rate cannot be drawn from the spin-transport measurements alone.  
 
 Our results also provide an alternative explanation to the observations of refs.\cite{dankert_electrical_2017, yan_two-dimensional_2016} where an enhanced spin-signal is observed when the TMD does not conduct. At this point, $E_{\text{F}}$ in graphene is shifted to the hole doped regime. Due to partial encapsulation of graphene via the TMD in refs.\cite{dankert_electrical_2017, yan_two-dimensional_2016}, the encapsulated and non-encapsulated  regions have different spin-transport properties, and the net spin-relaxation rate is dominated by the spin-relaxation at the graphene-TMD interface. It is reflected in a reduced value of $\Delta R_{\text{NL}}$ and $\tau_{\text{s}}$, coinciding with the conducting-state of the TMD for the electron-doped regime in graphene. Therefore, based on our results, we argue that it is the modulation of the spin-orbit coupling strength than the spin-absorption which changes the spin-relaxation time, leading to the same results.   
 \begin{figure*}[]
\includegraphics[]{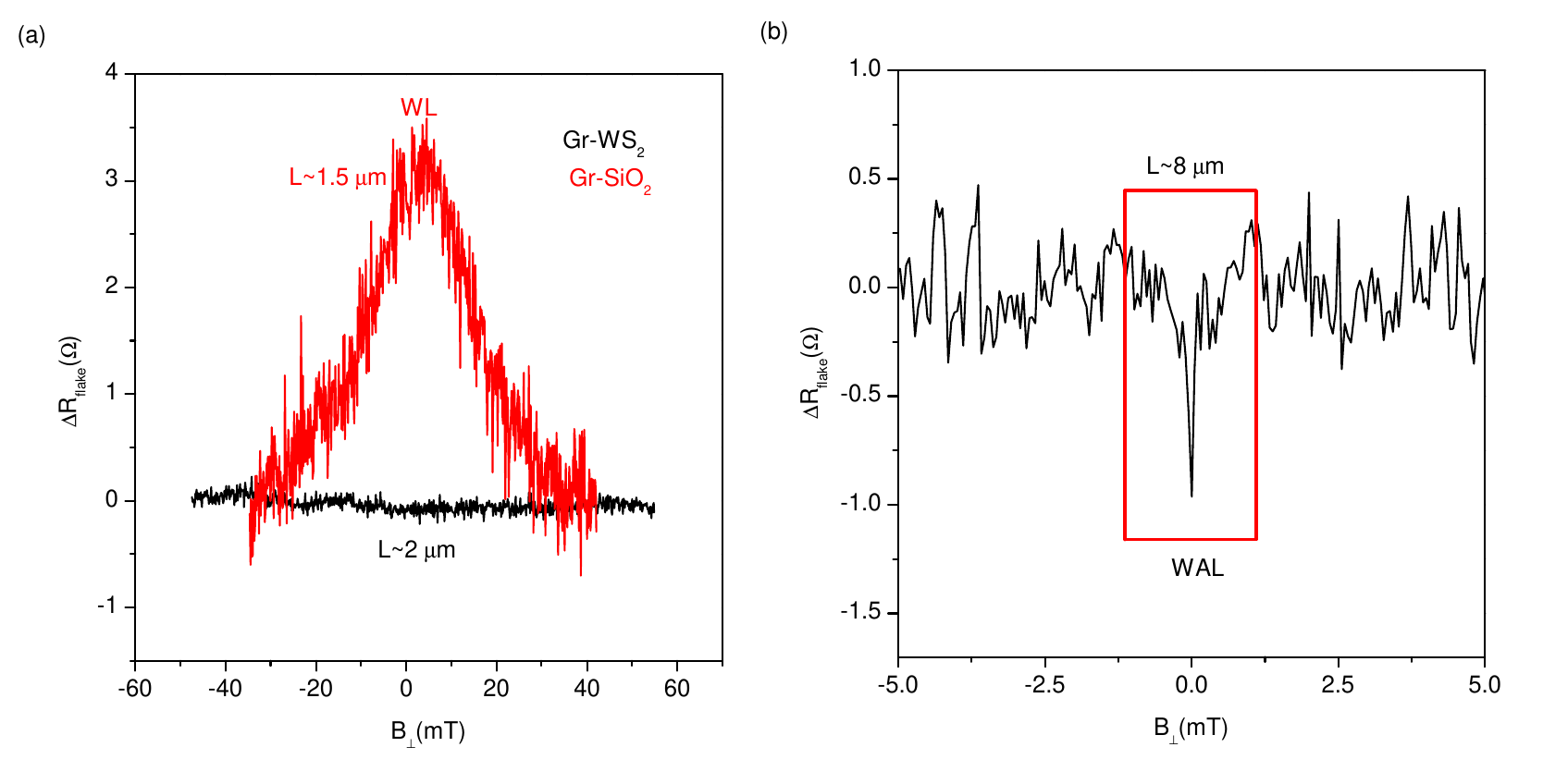}
\caption{\label{fig4}(a) A WL signal for ref B flake on a SiO$_2$  substrate (red) and no WL/WAL signature was detected for graphene-on-WS$_2$ (region-I of stack A). (b) A  narrow WAL signature in the encapsulated region was detected due to more spatial averaging  in a longer region (region-II of stack A). All the data shown here is taken at 4 K.} 
\end{figure*}

Alternatively, in order to confirm the presence of a substrate induced SOC in graphene, we perform the quantum magnetoresistance measurements in graphene in the electron-doped regime at 4 K, using the local four-probe geometry. Here we measure the flake resistance as a function of an out-of plane magnetic field with several averaging operations, in order to suppress the universal conductance fluctuations (UCF)  in the sample resistance at low temperatures \cite{lundeberg_defect-mediated_2013}. 
First, we measure the MR of the reference graphene-flake on SiO$_2$ (ref B) substrate at 4 K. Here we see a weak-localization (WL) signature (Fig.~\ref{fig4}(a)). A WL signature appears at low magnetic fields due to a suppressed back-scattering of electrons \cite{lundeberg_defect-mediated_2013}. A broad WL signal is probably due to the low mobility of graphene-on-SiO$_2$. \cite{yang_tunable_2016,lundeberg_defect-mediated_2013, wang_origin_2016}. However, for graphene-on-WS$_2$ (region-I, stack A) under the same measurement conditions, we do not observe any signature of the weak localization. For graphene-on-WS$_2$ we have even three times higher mobility than the reference sample which should help in observing a WL peak at a small range of the magnetic field \cite{wang_origin_2016}. The absence of the WL signal in graphene-on-WS$_2$ indicates the emergence of a competing behavior, for example due to the weak anti-localization effect. In fact, when we measure the MR for a longer graphene-channel of length $\sim$ 8 $\mu$m, including the encapsulated region, we observe a clear WAL signature (Fig.~\ref{fig4}(b)), which could be due to more spatial averaging of the signal in a longer graphene-channel. The observation of the WAL signature in the WS$_2$ supported single layer graphene confirms the existence of an enhanced SOC in graphene \cite{wang_origin_2016,wang_strong_2015}. 
 
 \section{conclusions}
In conclusion, we study the effect of a TMD (WS$_2$) substrate induced SOC in graphene via pure spin-transport measurements.  In spin-valve measurements for a broad carrier density range and independent of the conducting state of WS$_2$, we observe a constant spin-signal, and unambiguously show that the spin-absorption process is not the dominant mechanism limiting the spin-relaxation time in graphene on a WS$_2$ substrate. The proximity induced SOC reflects in broad Hanle curves with $\tau_{\text{s}} \sim$ 10-14 ps in the electron doped regime. Via the top-gate voltage application in the encapsulated region, we measure $\tau_{\text{s}} \sim$ 40-80 ps in the hole-doped regime, implying a reduced SOC strength. We also confirm the signature of the proximity induced SOC in graphene via WAL measurements. For both electron and hole regimes, we observe the DP-type spin-relaxation mechanism. The presence of the DP-type behavior is more (less) pronounced for the hole (electron) regime due to a higher (lower) $\tau_{\text{s}}$. We also demonstrate the modification of $\tau_{\text{s}}$ as a function of an out-of-plane electric field in the hBN-encapsulated region which suggests the control of in-plane Rashba SOC via the electrical gating.  In future experiments, in order to realize more effective control of electric field on $\tau_{\text{s}}$, the single layer graphene can be replaced by a bilayer graphene \cite{khoo_-demand_2017,gmitra_proximity_2017}. To enhance the spin-signal magnitude, a bilayer hBN tunnel barrier \cite{gurram_bias_2017} with a high spin-injection-detection efficiency can also be used. 
 
 Summarizing our results, we for the first time, unambiguously demonstrate the effect of the proximity induced SOC in graphene on a semi-conducting WS$_2$ substrate with high intrinsic SOC via pure spin-transport measurements, opening a new avenue for high mobility spintronic devices with enhanced spin-orbit strength. A gate controllable SOC and thus the modulation of $\tau_{\text{s}}$ almost by an order of magnitude in our graphene/WS$_2$ heterostructure paves a way for realizing the future spin-transistors. 
    
%
 \section{Acknowledgements}
We acknowledge  J. G. Holstein, H. M. de Roosz, T. Schouten and H. Adema for their technical assistance. We are extremely thankful to M. Gurram for the scientific discussion and his help during the sample preparation and measurements. This research work was funded from the European Union's Horizon 2020 research and innovation programme (grant no.696656) and supported by the Zernike Institute for Advanced Materials and the Netherlands Organization for Scientific Research (NWO).


%

\end{document}